\documentclass[useAMS,usenatbib]{mn2e}
\usepackage{amssymb}
\usepackage{epsfig}
\newcommand{\epshat}{{\widehat \epsilon}}
\newcommand{\mhat}{{\widehat m}}
\newcommand{\zhat}{{\widehat z}}
\newcommand{\Rhat}{{\widehat R}_{CV}}
\newcommand{\Rbar}{\bar{{\widehat R}}_{CV}}

\begin{document}

\title[Photometric Redshift Estimation Using SCA]{Photometric Redshift Estimation Using Spectral Connectivity Analysis}
\author[P. E. Freeman et al.]{P.~E.~Freeman$^1$\thanks{E-mail: pfreeman@cmu.edu}, J.~A.~Newman$^2$, A.~B.~Lee$^1$, J.~W.~Richards$^1$, C.~M.~Schafer$^1$\\
$^1$Department of Statistics, Carnegie Mellon University, 5000 Forbes Avenue, Pittsburgh, PA 15213\\
$^2$Department of Physics and Astronomy, University of Pittsburgh, 3941 O'Hara Street, Pittsburgh, PA 15260}

\date{12 April 2009}

\pagerange{\pageref{firstpage}--\pageref{lastpage}} \pubyear{2009}

\maketitle

\label{firstpage}

\begin{abstract}
The development of fast and accurate methods of photometric redshift estimation
is a vital step towards being able to fully utilize the data of
next-generation surveys within precision cosmology.  
In this paper we apply a specific approach to 
{\it spectral connectivity analysis} (SCA;
\citealt{LeeWasserman2009}) called diffusion map. 
SCA is a class of non-linear techniques
for transforming observed data (e.g., photometric
colours for each galaxy, where the data 
lie on a complex subset of $p$-dimensional
space) to a simpler, more natural coordinate system wherein we apply
regression to make redshift predictions.  
In previous applications of SCA to other
astronomical problems (\citealt{Richards09a}, \citealt{Richards09b}),
we demonstrate its superiority vis-a-vis Principal Components Analysis
(PCA), a standard linear technique for transforming data.
As SCA relies
upon eigen-decomposition, our training set size is limited to $\la$ 10$^4$
galaxies; we use the Nystr\"om extension to quickly estimate 
diffusion coordinates for objects
not in the training set.  We apply our method to 350,738 SDSS main sample
galaxies, 29,816 SDSS luminous red galaxies, and 5,223 galaxies from DEEP2
with CFHTLS $ugriz$ photometry.  For all three datasets,
we achieve prediction accuracies on par with previous analyses, and find 
that use of the Nystr\"om extension 
leads to a negligible loss of prediction accuracy relative
to that achieved with the training sets.
As in some previous analyses (e.g., \citealt{Collister04}, \citealt{Ball08}),
we observe
that our predictions are generally too high (low) in the low (high)
redshift regimes.
We demonstrate that this is a manifestation of attenuation
bias, wherein measurement error (i.e., uncertainty in diffusion coordinates
due to uncertainty in the measured fluxes/magnitudes) reduces the slope
of the best-fit regression line.  Mitigation of this bias is necessary if
we are to use photometric redshift estimates produced by computationally
efficient empirical methods in precision cosmology.
\end{abstract}


\section{Introduction}

The accurate estimation of redshifts from photometric data is a key
component to fulfilling the promise of next-generation cosmological surveys.
For instance, photometry to $R \sim$ 30 is expected for 
billions of galaxies from the Large Synoptic Survey Telescope (LSST;
\citealt{Ivezic08}) alone; compare this to, e.g., the $\sim$ 10$^5$ spectra
collected to a depth $R \sim$ 24 by the DEEP2 Galaxy Redshift Survey
(\citealt{Davis03}, \citealt{Davis07}).  It is clear that 
redshift-dependent analyses of galaxies that aim to undercover
signatures of, e.g., weak lensing or baryon acoustic oscillations 
in imaging data will require the use of photometric redshifts.
 
Redshifts estimated via, e.g., $ugriz$ photometry will necessarily lack
the precision of those that are spectroscopically derived due to noise, 
outliers, weak spectral features, and incomplete spectral energy
distribution (SED) templates.  For this reason, one major goal of
photometric redshift estimation is to generate accurate ensembles of
estimates, i.e., to have the mean redshift within a photometric redshift
bin be an accurate estimator of true redshift (see, e.g., 
\citealt{Albrecht06}, \citealt{Ma06}).
Such ensembles are typically generated via one of two
methods: either template fitting, wherein redshifted SED templates are 
generally
compared to a given vector of magnitudes with the goal of minimizing the
$\chi^2$ statistic or maximizing the likelihood
(e.g., \citealt{FernandezSoto99}, \citealt{Benitez00},
\citealt{Feldmann06}),
or empirical methods, wherein one
uses photometry from a small collection of objects with 
spectroscopically confirmed redshifts to train a model relating
photometric colours to redshifts (e.g., \citealt{Connolly95},
\citealt{Vanzella04}, \citealt{Collister04}, \citealt{Budavari05}, 
\citealt{Ball07}, \citealt{Ball08}, \citealt{Oyaizu08}).
Some combine the two approaches (e.g., \citealt{Ilbert06}, 
\citealt{Ilbert09}), while others propose folding in information
beyond photometric colours (e.g., \citealt{Collister04},
\citealt{Ball04}, \citealt{Wray08}, \citealt{Newman08}).

In this paper we propose a new empirical method for photometric
redshift estimation based on the diffusion map 
(\citealt{Coifman:Lafon:06}, \citealt{LafonLee2006}),
which is an approach to {\it spectral connectivity analysis (SCA)}.
SCA is a suite of established non-linear eigen-techniques\footnote{
The name SCA is applied to these eigen-techniques
by \cite{LeeWasserman2009}, who study their statistical properties.}
that capture the underlying geometry of data by
propagating local neighborhood information through a Markov process.
SCA thus allows one to find a natural coordinate system for data 
such as photometric colours whose original parametrization is not 
amenable to available statistical techniques.  
In \cite{Richards09a} and \cite{Richards09b}, we apply the diffusion map
to two different astronomical problems.  In \cite{Richards09a},
we develop a framework combining diffusion map and adaptive linear
regression and apply it to SDSS spectroscopic data, demonstrating how
it may be used to reduce the dimensionality of the data space and to 
predict, e.g., redshifts in a computationally efficient
manner. We also demonstrate the superiority of the diffusion map to
principal components analysis, a related, much more commonly 
used linear technique.  In \cite{Richards09b}, we utilize the diffusion
map and the K-means clustering algorithm to
determine optimal bases of simple stellar population spectra that we
use to estimate the star-formation histories of galaxies.

In {\S}\ref{sect:alg}, we review the basics of our diffusion map and
regression framework, and introduce a new component:
the application of the Nystr\"om 
extension (see, e.g., \citealt{NumRep}), a
computationally efficient and accurate technique
for estimating diffusion coordinates for new objects
given those of the training set.  
In {\S}\ref{sect:apply}, we apply our framework to
Sloan Digital Sky Survey data, specifically main sample galaxies
(MSGs) and luminous red galaxies (LRGs), and demonstrate that we
achieve accuracy on par with that of more computationally intensive
techniques.  We also apply our framework to data from the DEEP2
Galaxy Redshift Survey
that is matched to $ugriz$ photometry of the Canada-France-Hawaii
Telescope Legacy Survey (CFHTLS; \citealt{Gwyn08})
and demonstate that it provides
accurate estimation of redshifts to $z \approx 0.75$ given four colours alone.
We demonstrate that the bivariate distributions of 
photometric and spectroscopic redshifts for SDSS and DEEP2 are affected
by attenuation bias, the tendency of measurement error in the predictor
to reduce the slope of linear models.  Last, in {\S}\ref{sect:summary},
we summarize our results and discuss how we can extend our framework to
the high redshift regime where spectroscopic coverage will be incomplete.

\section{Algorithm}

\label{sect:alg}

\subsection{Diffusion map}

\label{sect:diff}

In this section we review the basics of diffusion map construction,
an approach to SCA.  
For more details, we refer the reader to \citet{Coifman:Lafon:06},
\citet{LafonLee2006}, and \citet{Richards09a}.  In
\citeauthor{Richards09a},
we compare and contrast the use of diffusion maps with a more
commonly utilized linear technique, principal components analysis,
and demonstrate the superiority of diffusion maps in predicting 
spectroscopic redshifts of SDSS data from the galaxy spectra.

Here,``spectral connectivity analysis'' refers to a class of
methods which utilize a local distance measure to ``connect''
similar observations. The eigenmodes (i.e., ``spectral decomposition'') 
of the rescaled matrix of similarities (see below for the definition
of this matrix) can reveal a natural coordinate system for data that
was absent in the original representation.  For instance, imagine
data in two dimensions that to the eye clearly exhibit spiral 
structure (e.g., fig. 1 of \citealt{Richards09a}).
For such data, the Euclidean distance between
data points $\bmath{x}$ and $\bmath{y}$ 
would not be an optimal description of the 
`true' distance between them along the spiral.  
Diffusion map is a leading example of an approach to SCA.
In the diffusion
map framework, the `true'
distance is estimated via a fictive diffusion process over the data,
with one proceeding from $\bmath{x}$ to 
$\bmath{y}$ via a random walk along the spiral.

We construct diffusion maps as follows.

We define a similarity measure $s(\bmath{x},\bmath{y})$ that quantitatively 
relates two data points $\bmath{x}$ and $\bmath{y}$.
In this work, a data `point' is a vector of colours \{$c_1,\dots,c_p$\} 
of length $p$ for a single galaxy, 
and the similarity measure that we apply is the Euclidean distance
\begin{eqnarray}
s(\bmath{x},\bmath{y})~=~\sqrt{\sum_{i=1}^p \left( c_{\bmath{x},i} - c_{\bmath{y},i} \right)^2 } \nonumber \,.
\end{eqnarray}
A key feature of SCA is that 
the choice of $s(\bmath{x},\bmath{y})$ is not crucial, as it is often simple to 
determine whether or not two data points are `similar.'

We remove extreme outliers from our dataset, not
because of their effect on diffusion map construction
(a hallmark of the diffusion map is its robustness in the presence
of outliers), but rather because they can bias the coefficients of the
linear regression model (see {\S}\ref{sect:regress}) and because we find
that individual predictions made for these objects are highly inaccurate.
We compute the empirical distributions of Euclidean distances in colour space
from each object to its $n^{\rm th}$ nearest neighbor, where 
$n \in [1,10]$.  These 
distributions are well-described as exponential, with estimated
mean and standard deviation 
${\hat \mu}_n = {\hat \sigma}_n = {\tilde x}_n/\log(2)$ 
for median value ${\tilde x}_n$.
We exclude all data whose $n^{\rm th}$ nearest neighbor is at a distance
$> {\hat \mu}_n + 5{\hat \sigma}_n = 6{\hat \sigma}_n$, for any value
of $n \in [1,10]$.  We find that 
$\approx$ 80\% of extreme outliers are removed with the first 
nearest-neighbor cut alone, with the fraction of those removed
falling as $n$ increases.

With outliers removed, we 
construct a weighted graph where the nodes are the observed data points:
\begin{eqnarray}
w(\bmath{x},\bmath{y})~=~\exp\left( -\frac{s(\bmath{x},\bmath{y})^2}{\epsilon} \right) \,,
\label{eqn:weighted}
\end{eqnarray}
where $\epsilon$ is a tuning parameter that should be small enough that
$w(\bmath{x},\bmath{y}) \approx 0$ unless $\bmath{x}$ and $\bmath{y}$ are 
similar, but large enough such that the graph is fully connected.  
(We discuss how we estimate $\epsilon$ in {\S}\ref{sect:regress}.)
The probability of stepping
from $\bmath{x}$ to $\bmath{y}$ in
one step is 
$p_1(\bmath{x},\bmath{y})~=~w(\bmath{x},\bmath{y})/\sum_z w(\bmath{x},\bmath{z})$.
We store the one-step probabilities between all $n$ data points in 
an $n \times n$ matrix \textbfss{P}; then, by the theory of Markov chains, the
probability of stepping from $\bmath{x}$ to $\bmath{y}$ in $t$ steps is given by
the element $p_t(\bmath{x},\bmath{y})$ of the matrix \textbfss{P}$^t$.  
The diffusion distance
between $\bmath{x}$ and $\bmath{y}$ at time $t$ is defined as
\begin{eqnarray}
D_t^2(\bmath{x},\bmath{y})~=~\sum_{j=1}^{\infty} \lambda_j^{2t}(\bpsi_j(\bmath{x})-\bpsi_j(\bmath{y}))^2 \nonumber \,,
\end{eqnarray}
where $\bpsi_j$ and 
$\lambda_j$ represent eigenvectors and eigenvalues of 
\textbfss{P}, respectively.  By retaining the $m$ eigenmodes corresponding
to the $m$ largest nontrivial eigenvalues and by introducing the diffusion map
\begin{equation}
\mathbf{\Psi}_t: \bmath{x} \mapsto [\lambda_1^t\bpsi_1(\bmath{x}), \lambda_2^t\bpsi_2(\bmath{x}),\cdots,\lambda_m^t\bpsi_m(\bmath{x})]
\label{eqn:diffusion_map}
\end{equation}
from $\mathbb{R}^p$ to $\mathbb{R}^m$, we have that
\begin{eqnarray}
D^2_t(\bmath{x},\bmath{y})~\simeq~\sum_{j=1}^m \lambda_j^{2t}(\bpsi_j(\bmath{x})-\bpsi_j(\bmath{y}))^2~=~||\mathbf{\Psi}_t(\bmath{x}) - \mathbf{\Psi}_t(\bmath{y})||^2 \,, \nonumber
\end{eqnarray}
i.e., the Euclidean distance in the $m$-dimensional embedding defined 
by equation~\ref{eqn:diffusion_map} approximates diffusion distance.
(We discuss how we estimate $m$ in {\S}\ref{sect:regress},
and show that, in this work, the choice of $t$ is unimportant.)
We stress that the diffusion map reparametrizes the data into a
coordinate system that reflects the connectivity of the data,
and does not necessarily affect dimension reduction.
If the original parametrization in
$\mathbb{R}^p$ is sufficiently complex, then it may be the case that $m \gg p$.

\subsection{Regression}

\label{sect:regress}

As in \citet{Richards09a}, we perform linear regression 
to predict the function
$z = r(\mathbf{\Psi}_t)$, where $z$ is true redshift and $\mathbf{\Psi}_t$ is 
a vector of diffusion coordinates in $\mathbb{R}^m$, representing
a vector of photometric colours $\bmath{x}$ in $\mathbb{R}^p$:
\begin{eqnarray}
{\widehat r}(\mathbf{\Psi}_t)~&=&~\mathbf{\Psi}_t {\widehat \beta}~=~\sum_{j=1}^m {\widehat \beta}_j \mathbf{\Psi}_{t,j}(\bmath{x}) \nonumber \\
&=&~\sum_{j=1}^m {\widehat \beta}_j \lambda_j^t \bpsi_j(\bmath{x})~=~\sum_{j=1}^m {\widehat \beta}_j' \bpsi_j(\bmath{x}) \nonumber
\end{eqnarray}
We see that the choice of the parameter $t$ is unimportant,
as changing it simply leads to a rescaling in ${\widehat \beta}_j$,
with no change in ${\widehat \beta}_j'$.  We present relevant 
regression formulae in Appendix \ref{sect:regform}.

We determine optimal values of the tuning parameters 
($\epsilon$,$m$) by minimizing estimates of the prediction risk,
$R(\epsilon,m)~=~\mathbb{E}(L)$,
where $\mathbb{E}(L)$ is the expected value of a loss function $L$ over
all possible realizations of the data 
(one example of $L$ is the so-called $L_2$ loss function, which
is simply the mean-squared error of the fit; 
see, e.g., \citealt{Wasserman2006} for
a discussion of this and other topics introduced below).
$R$ quantifies the `bias-variance' tradeoff: too much smoothing ($m$ too low)
yields prediction estimators with low variance and high bias, while too
little smoothing ($m$ too high) yields estimators with high variance and low
bias.  Using the full data set to estimate $R$
underestimates the error and leads to a best-fit
model with high bias, thus we apply $10$-fold cross-validation (CV). 
The data are partitioned into 10 blocks of (approximately) equal size.
We regress upon the data in nine of the blocks and use the best-fit
regression model to predict the responses $\zhat_i$ for the data 
in the tenth block.  
(We note that for algorithmic consistency we use the Nystr\"om extension
to estimate the diffusion coordinates of the data in the tenth block;
see {\S}\ref{sect:nystrom}.)
The process is repeated 10 times, for different block combinations,
so that predictions are generated for each datum.
The individual predictions are combined into an overall risk estimate
\begin{eqnarray}
\Rhat(\epsilon,m)~=~\sqrt{\frac{1}{n} \sum_i \delta_i^2}~=~\sqrt{\frac{1}{n} \sum_i \left(\frac{\vert \zhat_i - Z_i \vert }{1 + Z_i}\right)^2} \,. \label{eqn:rcv}
\end{eqnarray}
where we apply the redshift-corrected rms dispersion
as our loss function.
$Z_i$ is the estimated spectroscopic redshift for object $i$.
(We capitalize to underscore the fact that the spectroscopic redshift 
is a random variable not necessarily equal to the true redshift $z_i$.)
To ensure robustness, for each set of tuning parameters $(\epsilon,m)$, 
we compute the mean $\Rbar(\epsilon,m)$ of 10 estimates of $\Rhat(\epsilon,m)$,
and select those values of $(\epsilon,m)$ such that $\Rbar(\epsilon,m)$ is
minimized, i.e., $(\epshat,\mhat) = {\rm arg~min} \:\Rbar(\epsilon,m)$.

\subsection{Diffusion coordinate estimation via the Nystr\"om extension}

\label{sect:nystrom}

The computation of diffusion coordinates (equation~\ref{eqn:diffusion_map}) 
relies upon eigen-decomposition, which is computationally intractable for
datasets of $\ga$ 10$^4$ galaxies.  (However, see \citealt{Budavari09}, who
propose an incremental methodology for computing eigenvectors.)
Photometric datasets can, of
course, be much larger, and thus we require a computationally efficient
scheme for estimating eigenvectors for new galaxies given those
computed for a small set of galaxies used to train
the regression model.  A standard method in applied mathematics for
`extending' a set of eigenvectors is the Nystr\"om extension.

The implementation
is simple: determine the distance in colour space from each new galaxy
to its nearest neighbors in the training set, then take
a weighted average of those neighbors' eigenvectors.  Let \textbfss{X} 
represent the $n \times k$ matrix containing the colour data of the 
training set, where $n$ and $k$ are the number of
objects and colours, respectively.  Let \textbfss{X'} represent
a similar $n' \times k$ matrix containing colour data for
$n'$ objects in the validation set.  The first step
of the Nystr\"om extension is to compute the $n' \times n$ weight matrix
\textbfss{W}, with elements equivalent to those shown in 
equation~\ref{eqn:weighted} above (except that there, $\bmath{x}$ and $\bmath{y}$ are
both members of the training set, while here, $\bmath{x}$ is a new point while
$\bmath{y}$ belongs to the training set).
We assume the same value $\epshat$ as was selected during diffusion map
construction; since the training set is a random sample of galaxies from 
our original set, we expect the validation set to be sampled from the
same underlying probability distribution.
We row-normalize \textbfss{W} by dividing by each element in
row $i'$ by $\rho_{i'} = \sum_i W_{i',i}$.

Let $\mathbf{\Psi}$ be the $n \times m$ matrix of eigenvectors with 
corresponding vector of eigenvalues $\lambda$.  To estimate the eigenvectors
for the new galaxies, we compute the $n' \times m$ matrix 
$\mathbf{\Psi}'$:
\begin{eqnarray}
\mathbf{\Psi}'~=~\mathbfss{W} \mathbf{\Psi} \mathbf{\Lambda} \,,
\label{eqn:nys}
\end{eqnarray}
where $\mathbf{\Lambda}$ is a $m \times m$
diagonal matrix with entries $1/\lambda_i$.  Then the redshift
predictions for the $n'$ objects are 
$\bmath{\zhat} = \mathbf{\Psi}' \bmath{\widehat \beta}$, 
where $\bmath{\widehat \beta}$
are the linear regression coefficients generated for the original training set.

\section{Application to SDSS and DEEP2 datasets}

\label{sect:apply}

\subsection{SDSS spectroscopic data}

In this work,
we use the Princeton/MIT reductions of SDSS spectroscopic data\footnote{
See \tt http://spectro.princeton.edu}.  
Features of these data include the so-called
`uber-calibration' of $ugriz$ magnitudes in six magnitude systems
(\citealt{Pad08}).  To facilitate a direct comparison of our results with
those of \citealt{Ball08}, we utilize colours, i.e.,
differences between the magnitudes measured in different bands determined
in each of four magnitude systems:
{\it psf}, {\it fiber}, {\it petrosian}, and {\it model}.  Thus the
colour data occupy a $p$ = 16 dimensional space.

The necessary data are contained in the files {\tt spAll-$<$rel$>$.fits},
where {\tt $<$rel$>$} = EDR and DR1$-$DR6.
We extract data from all publicly available plates for which
{\tt PROGNAME} = `main' and {\tt PLATEQUALITY} = `good,' keeping
1001 plates in all.  (We keep only one instance of each plate when
repeated observations are made, making the ad hoc choice to retain the
most recent observation.)
For each plate, we examine data for those fibers for
which {\tt CLASS} = `GALAXY,' {\tt Z} $>$ 0.01,
and {\tt ZWARNING} = 0.  For each of these fibers, we apply 
extinction corrections \{$A$\} (from column {\tt EXTINCTION}) 
to the set of fluxes $\{F\}$
and the set of estimated standard errors $\{s_F\}$ (\citealt{Fink04}):
\begin{eqnarray}
F'~&=&~10^{0.4A}F \nonumber \\
s_{F'}~&=&~\frac{s_F}{\sqrt{10^{-0.8A}}} \nonumber \,.
\end{eqnarray}
If for any object, one or more elements of the 
set $\{F'\} < 0$, we exclude the object from analysis.
The flux units are nanomaggies; the conversion from $F'$ to magnitude $m'$
is $m' = 22.5 - 2.5 \log_{10} F'$, while the conversion to colours is
$c_{i-j}' = 2.5 \log_{10}(F_j'/F_i')$.  

The final number of galaxies in our sample is 417,224.

\subsubsection{Main sample galaxies}

From our data sample, we extract those 360,122 galaxies with Petrosian
$r$-band magnitude $< 17.77$ (or $F_{\rm Petro}^R >$ 77.983; 
\citealt{Strauss02}).  
This is our main sample galaxy or MSG sample.
We randomly select 10,000 galaxies from this sample
to train our regression model.  Application of the outlier-removal algorithm
described in {\S}\ref{sect:regress} leads to the removal of 251 galaxies from
this set.  
The application of the algorithm outlined in {\S}{\S}2.1-2 yields tuning
parameter estimates ($\epshat,\mhat$) = (0.05,150), i.e., in order for a
linear model to be appropriate, the 16-dimensional colour data is reparametrized
into 150-dimensional space.  

As each object's eigenvector estimates are independent of those for
other objects, we apply the Nystr\"om extension 
to validation set objects one plate at a time,
then concatenate the resulting predictions.  We determine which members of
the validation set are 5$\sigma$ outliers relative to the members of the
training set, and compute the value of $\Rhat$ with those objects
excluded.  (Not excluding these outliers, which lie too far from the
training set in colour space for their diffusion coordinates 
to be estimated accurately,
results in $\Rhat$ rising from $\approx$ 0.02 to 0.56.)
Out of 350,122 objects in the validation set, we exclude
9,133; the percentage of outliers is 2.61\%.  This is consistent
with the 2.51\% rate of outliers in the training set.

\begin{table}
\caption{Parameters of optimal regression}
\label{symbols}
\begin{tabular}{lccccc}
\hline
Dataset & $(\epshat,\mhat)$ & ${\widehat R}_{CV}$ & $\eta$ (\%) & $n$ & $n_{\rm out}$ \\
\hline
MSG-T & (0.05,150) & 0.0206 & 0.010 & 9,749 & 251 \\
          &            & (0.0231) \\
MSG-V & & 0.0211 & 0.018 & 340,989 & 9,384 \\
 &            & (0.0240) \\
LRG-T & (0.012,200) & 0.0189  & 0.010 & 9,734 & 266 \\
          &            & (0.0258) \\
LRG-V & & 0.0195 & 0.034 & 20,082 & 884 \\
          &            & (0.0270) \\
DEEP2-T & (0.002,850) & 0.0507 & 1.67 & 5,223 & 304 \\
 &            & (0.1063) \\
DEEP2-T & (0.002,1050) & 0.0539 & 2.14 & 6,067 & 351 \\
       &            & (0.1123) \\
\hline
\end{tabular}

\medskip

In the column `Dataset,' T = training set and V = validation set.
$\eta$ is the rate of catastrophic failures (i.e., the rate at which
$\delta > 0.15$),
$n$ is the number of galaxies used in analysis after outlier removal, and
$n_{out}$ is the number of 5$\sigma$ outliers removed from sample.
The number (outside/inside) the parantheses in column ${\widehat R}_{CV}$
(includes/does not include) normalization by $(1+Z)$.
$u$-band data are excluded from LRG analyses.
For {\em DEEP2-T}, the first and second rows represent analyses of
objects for which {\tt ZQUALITY} = 4 and {\tt ZQUALITY} $\geq$ 3, respectively.
\label{tab:results}
\end{table}

\begin{figure}
\includegraphics[width=84mm]{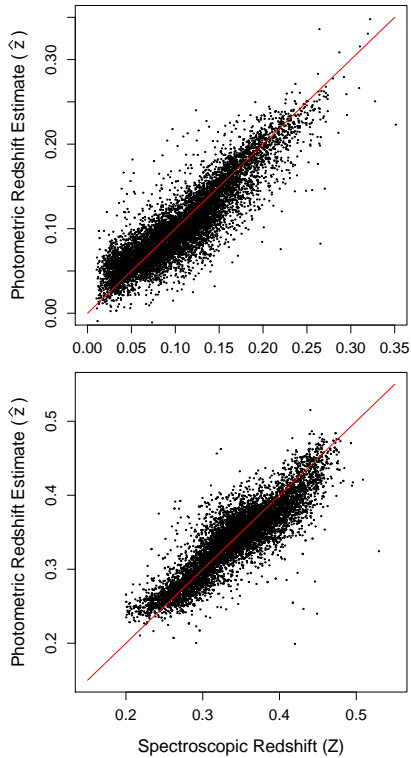}
\caption{Top: predictions for 10,000 randomly selected objects in 
the MSG validation set, for $(\epshat,\mhat)$ = (0.05,150).
Bottom: same as top, for the LRG validation set, with
$(\epshat,\mhat)$ = (0.012,200).
In both cases, we remove 5$\sigma$ outliers from the sample
prior to plotting, thus the actual number of plotted points 
is 9,740 (top) and 9,579 (bottom).
}
\label{fig:sdss}
\end{figure}

We show our results in Table \ref{tab:results} and the top panel
of Fig.~\ref{fig:sdss}, in which we display
predictions for 10,000 randomly chosen objects of the validation set.
The accuracy of prediction via the Nystr\"om extension versus directly
fitting a linear regression model to the diffusion map coordinates
of the data is indicated in
Table \ref{tab:results}.  We find that $\Rhat$ increases by
2.4\% from 0.0206 to 0.0211, with 
catastrophic failure rate $\eta$ increasing but still small.
(Here, a catastrophic failure for object $i$ is defined as 
$\delta_i >$ 0.15; see equation \ref{eqn:rcv} and, e.g., \citealt{Ilbert06}.)
The small degradation in accuracy is more than balanced by computational 
speed; our naive implementation allowed
extension to 350,373 galaxies in $\sim$ 10 CPU hours on 
a single GHz processor, a computation time that will be markedly
reduced in future implementations of the algorithm.
$\Rhat$ = 0.0211 (0.0240 without normalization by $1+Z$)
compares favorably with a myriad of other 
analyses of MSG data (see, e.g.,
\citeauthor{Ball08}, who obtain $\sigma$ = 0.0207 without $1+Z$ normalization,
and references therein), and
the empirical bivariate distribution of $(\zhat,Z)$ is 
visually indistinguishable from those of, e.g., \citeauthor{Ball08}
and \cite{Collister04}.

\begin{figure}
\includegraphics[width=84mm]{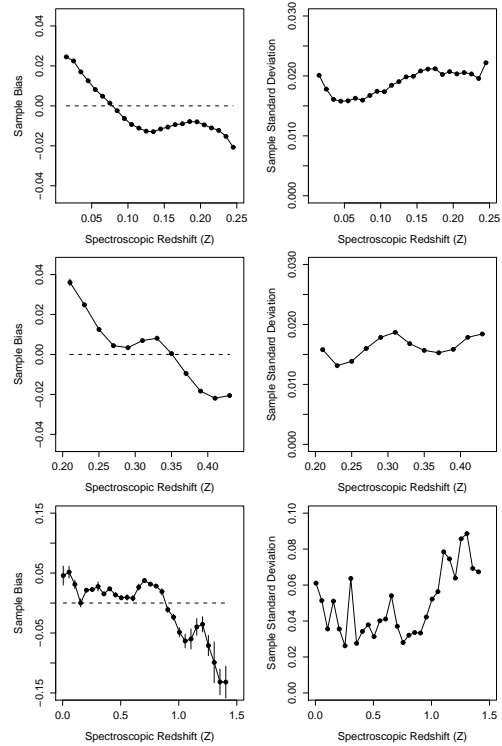}
\caption{
Top Left: estimated bias $\zhat - Z$ for MSG redshift estimates $\zhat$, 
computed in bins of width $\Delta Z$ = 0.01 in the range $Z \in [0.01,0.25]$.
Top Right: estimated standard deviation for MSG redshift estimates
(normalized by $1+Z$).  
Middle Left and Right: same as top left and right,
except for LRG redshift estimates in bins of width $\Delta Z$ = 0.02 
in the range $Z \in [0.20,0.44]$.
Bottom Left and Right: same as top left and right,
except for DEEP2 redshift estimates ({\tt ZQUALITY} = 4)
in bins of width $\Delta Z$ = 0.05 
in the range $Z \in [0.0,1.5]$.
}
\label{fig:bias}
\end{figure}

\begin{figure}
\includegraphics[width=84mm]{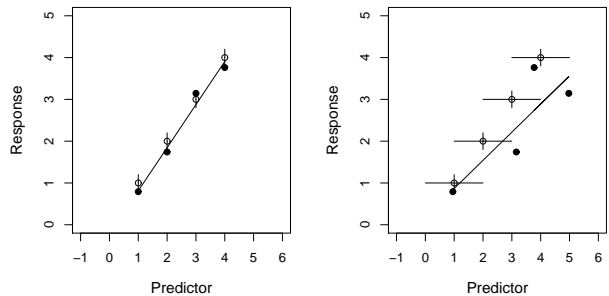}
\caption{
Simple demonstration of the effect of attenuation bias on linear
regression.  Left: example of linear regression fit to data with
no measurement error in the predictor and with response $Y \sim N(x,0.04)$,
where $x = \{1,2,3,4\}$, i.e.,
each value of $Y$ is sampled from a Gaussian distribution with mean
$x$ and variance 0.04.
The black dots indicate the observed data,
while the open circles show the true $(x,y)$ values.
Right: same as left, but with measurement error applied to the
predictor: $X \sim N(x,1)$.
The effect of this measurement error is to reduce the slope of the 
regression line, on average.  The mean reduction in slope for this
toy example is 0.25 (from 1 to 0.75), as estimated via 10,000 simulations.
}
\label{fig:atten}
\end{figure}

\begin{figure}
\includegraphics[width=84mm]{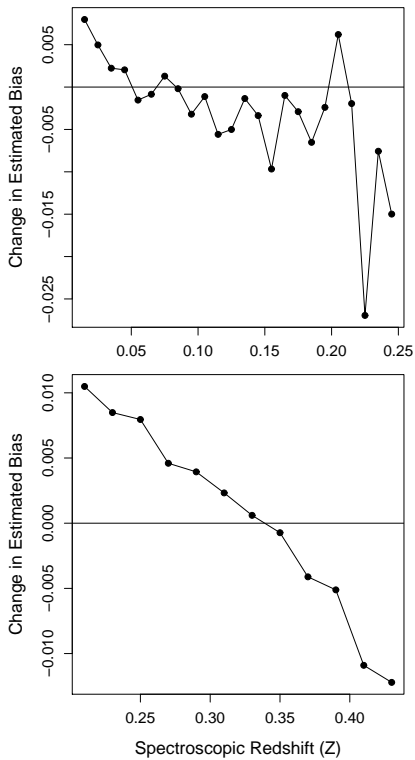}
\caption{
Top: change in the estimated bias $\zhat - Z$ induced by resampling
MSG training and validation set fluxes and refitting.  Because resampling
increases the measurement error (i.e., the error in the predictor, in this
case the diffusion coordinates), the slope of the regression line is
reduced, increasing overestimates of $\zhat$ at low $Z$ and underestimates
of $\zhat$ at high $Z$.  Bottom: same as top, for LRG datasets.
}
\label{fig:comp_bias}
\end{figure}

We determine estimator bias by binning
the predictions $\zhat$ as a function of $Z$, then
in each bin computing $\bar \zhat - Z$, with
$\bar \zhat$ being a 10\% trimmed mean. 
See the top left panel of Fig.~\ref{fig:bias}.
It is readily apparent that there is a downward slope in the bias
(i.e., redshifts are overestimated at low $Z$, and underestimated at
high $Z$).  This is not caused by model bias (a bias that one would
mitigate by adding complexity to the model, e.g., changing from linear
to quadratic regression), but rather by {\it attenuation bias}, 
in which measurement error 
(i.e., uncertainty in the predictor, in this case the diffusion
coordinates) reduces the slope of the regression line 
(see Fig. \ref{fig:atten}; see also, e.g., \citealt{Carroll06}).
To demonstrate that our data are affected by attenuation bias, we perform
a simple experiment.  First, we take the MSG
training set fluxes and resample
them according to the prescription given in Appendix \ref{sect:sample}.
This increases all measurement errors.  (To see this intuitively, imagine
sampling random variables $X \sim N(0,1)$, i.e.,
each value of $X$ is sampled from a Gaussian distribution with mean
0 and variance 1.
Then resample from the observed values $X$: $Y \sim N(X,1)$.  The
standard deviation of the resulting sample is now $\sqrt{2}$, i.e., the
error has been artificially increased by resampling.)
Then we resample fluxes for 1,000 randomly selected validation set
objects.  By doing each resampling (training set and validation set) 25
times, we build up a set of 625 predictions of $\zhat$ for each of the 1,000
selected objects.  Following the same prescription as above, we estimate
the bias; the top panel of Fig. \ref{fig:comp_bias} shows how for the MSG
dataset, increasing the measurement error via resampling leads to
a steepening of the bias slope, i.e., the effect of attenuation bias
is magnified.

There exist methods for correcting the bias in linear regression coefficient
estimation caused by additive, heteroscedastic (i.e., non-constant)
measurement errors of known magnitude that are
based on the {\tt SIMEX}, or simulation-extrapolation, algorithm 
(\citealt{Cook94}; see, e.g., \citealt{Carroll06} and references therein).
Indeed, one of the advantages to our approach is that the non-linearity
is in the reparametrization, not the fitted model. Hence, available methods
for correcting for measurement error could be utilized.
We are currently exploring the implementation of {\tt SIMEX}-based
methods in a computationally efficient manner, and we will 
present our results in a future publication.

\begin{figure}
\includegraphics[width=84mm]{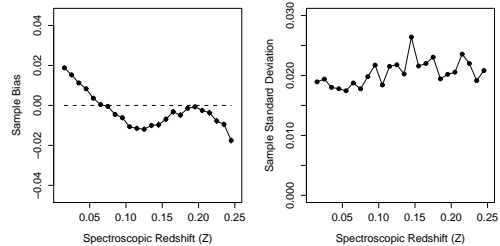}
\caption{
Left: estimated bias $\zhat - Z$ for MSG redshift estimates $\zhat$, 
computed in bins of width $\Delta Z$ = 0.01 in the range $Z \in [0.01,0.25]$,
for a 10,000-galaxy sample constructed so as to be uniform in $Z$.
Uniformity in $Z$ reduces the bias slightly (cf.~the top panel of
Fig.~\ref{fig:bias}).  This result indicates that measurement error is
the dominant cause of the bias.
Right: estimated standard deviation for MSG redshift estimates
(normalized by $1+Z$).  
}
\label{fig:bias_unif}
\end{figure}

While attenuation bias is caused by measurement error, its magnitude is
affected by the distribution of the predictors, i.e., the design.  
Expressions relating the design to the bias magnitude are highly problem
dependent.  In the
simplest, one-dimensional example of attenuation bias, the predictors are 
assumed to be normally distributed--$X \sim N(\mu_x,\sigma_x^2)$--and the
effect on the slope $\beta_1$ is to reduce its value:
$\hat{\beta_1} \rightarrow \lambda \beta_1$, where $\lambda = 
\sigma_x^2/(\sigma_x^2 + \sigma_u^2)$ and $\sigma_u$ is the measurement error.
The smaller the value of $\sigma_x^2$, the greater the effect upon the bias.
We mention this explicitly to underscore that analyzing samples for which
the predictors are, e.g., uniformly distributed may reduce the magnitude of
the bias magnitude but will not eliminate it since measurement error is still
present.  In Fig.~\ref{fig:bias_unif}, we show the estimated sample bias
as a function of $Z$ for a 10,000-galaxy sample constructed so as to be
uniform in $Z$ (though the
distribution of the predictors themselves--the diffusion coordinates--is not 
necessarily uniform).  Comparing these results with the top panels of
Fig.~\ref{fig:bias}, we find that uniformity in $Z$ reduces the bias slightly
(while also slightly increasing sample standard
deviation).  This indicates that measurement error is the dominant cause
of the observed bias.

Nonparametric estimators such as k-nearest neighbor
(kNN) and local polynomial regression are also affected by measurement error 
bias (whose mitigation is dubbed the ``deconvolution problem")
and design bias, and in addition by boundary bias (see, e.g., 
chapter 5 of \citealt{Wasserman2006} and 
chapter 12 of \citealt{Carroll06} and references therein).
Thus the similarity
of our bivariate distribution to that of, e.g., \citeauthor{Ball08}  (See
their fig.~6.  In this figure, we note slightly larger deviations from the 
$\hat{z} = Z$ locus at the endpoints than our bivariate distribution exhibits,
which may indicate boundary bias but also could be a result of the fact
that \citeauthor{Ball08} do not minimize risk and thus could be adopting
a solution with relatively higher bias and lower variance than our solution.)

In addition to estimator bias, we also examine the estimator variance, i.e.,
the width of the observed bivariate distribution 
(given as a function of $Z$ in the right column of Fig.~\ref{fig:bias}).
Contributing to the variance
is (a) model uncertainty, i.e., the standard deviation of the estimates
$\zhat$ (given by the square root of the diagonal elements of the matrix
given in equation~\ref{eqn:vy}); (b) uncertainty in the flux 
for each object;
and (c) intrinsic scatter, i.e., the
fact that the MSG sample does not necessarily contain a homogeneous set of
objects.  Model uncertainty contributes little to the observed scatter;
the mean, median, and standard deviation of 
the model uncertainties are $\la$ 10$^{-5}$.  Flux uncertainty enters
via attenuation bias; as flux errors increase, 
the linear regression slope flattens and acts to decrease the sample
variance within a redshift bin.  However, in our simple 
attenuation-bias demonstration we observe only negligible changes
in the sample variance.
Thus we conclude that the observed sample variance
is primarily due to intrinsic scatter,
and can only be reduced by introducing more data (cf.~\citealt{Ilbert09},
who achieve $\Rhat \la$ 0.01 by utilizing data from 30 bands in the
UV, optical, and IR regimes).

\subsubsection{Luminous red galaxies}

From our data sample, we extract those 30,700 galaxies for which 
$Z > 0.2$ and {\tt PRIMTARGET} = 32 ({\tt TARGETGALAXYRED}; \citealt{Eisen01}).
This is our luminous red galaxy or LRG sample.  As with the MSG training
set, we randomly select 10,000 galaxies and then remove outliers.  
Because the $u$ band data of high-redshift LRGs lacks constraining power
(as LRGs are faint in $u$ and thus the magnitudes are noisy),
we use only $griz$ fluxes in analyses
(so that $p$ = 12).  The training set contains 9,734 objects. 
Application of the algorithm outlined in {\S}{\S}2.1-2.2
yields tuning parameter estimates $(\epshat,\mhat)$ = (0.012,200).
The results of fitting are shown in Table \ref{tab:results} and
the bottom panel of Fig. \ref{fig:sdss}.  As in the case of the
MSG analysis, our value $\Rhat$ = 0.0195 (0.0270 without $1+Z$ normalization)
compares favorably with, e.g., \citealt{Ball08}, who achieve
$\sigma$ = 0.0242 (without $1+Z$ normalization), and references therein.
We find that the outlier rate is consistent from training
set to validation set (increasing from 2.7\% to 3.1\%), 
and that $\Rhat$ increases by only 3.1\%
when we use the Nystr\"om extension as opposed to directly fitting the data.
(Note that if we include the $u$ band, the estimate of $\epshat$ increases by
two orders of magnitude, indicating the scatter in colour space introduced
by non-constraining $u$-band data, although
$\Rhat$ itself only rises by $\approx$ 5\%.)
The LRG redshift predictions, like their MSG counterparts, are biased,
with a similar downward trend in the bias as a function of $Z$
(left middle panel, Fig.~\ref{fig:bias}).  We repeat
our simple resampling experiment with LRG data and find that the bias slope
increases upon resampling, demonstrating that attenuation bias also 
affects LRG data analysis (as expected; see Fig.~\ref{fig:comp_bias}).

\subsection{DEEP2/CFHTLS data}

The DEEP2 Galaxy Redshift Survey (\citealt{Davis03}, \citealt{Davis07}) 
studied both galaxy properties and large-scale structure
primarily at redshifts $0.7 \la z \la 1.4$,
in four fields of total area $\sim$ 3 square degrees.
DEEP2 targets are selected to have $R_{AB} \leq$ 24.1 using CFHT BRI
photometric data (\citealt{Coil04}).
In three of the four DEEP2 fields, colour cuts are used to select
$z >$ 0.7 objects for observation; however, in this paper we utilize
the DEEP2 sample in the Extended Groth Strip, for which no colour cuts
have been applied.  DEEP2 collected spectra typically covering the
wavelength range 6,500$-$9,100 \AA~for $>$ 50,000 objects.
From the survey we select the 6,552 galaxies for which single-system
$ugriz$ photometry
exists from the CFHT Legacy Survey
(field D3)\footnote{
See {\tt http://www4.cadc-ccda.hia-iha.nrc-cnrc.gc.ca/\\
community/CFHTLS-SG/docs/cfhtls.html} and \cite{Gwyn08}.
}
and for which
the DEEP2 {\tt ZQUALITY} flag is either 3 or 4 
($>$ 95\% or 99.5\% confidence that the redshift is correct, respectively).
Thus the dimensionality of colour-space for these data is $p$ = 4.
We further remove data for which the redshift error, or any magnitude or 
magnitude error, is not provided, leaving 6,418 galaxies; after outlier
removal, the final sample size is 6,067.  If we restrict ourselves to
data for which {\tt ZQUALITY} = 4, the sample size is 5,223.

\begin{figure}
\includegraphics[width=84mm]{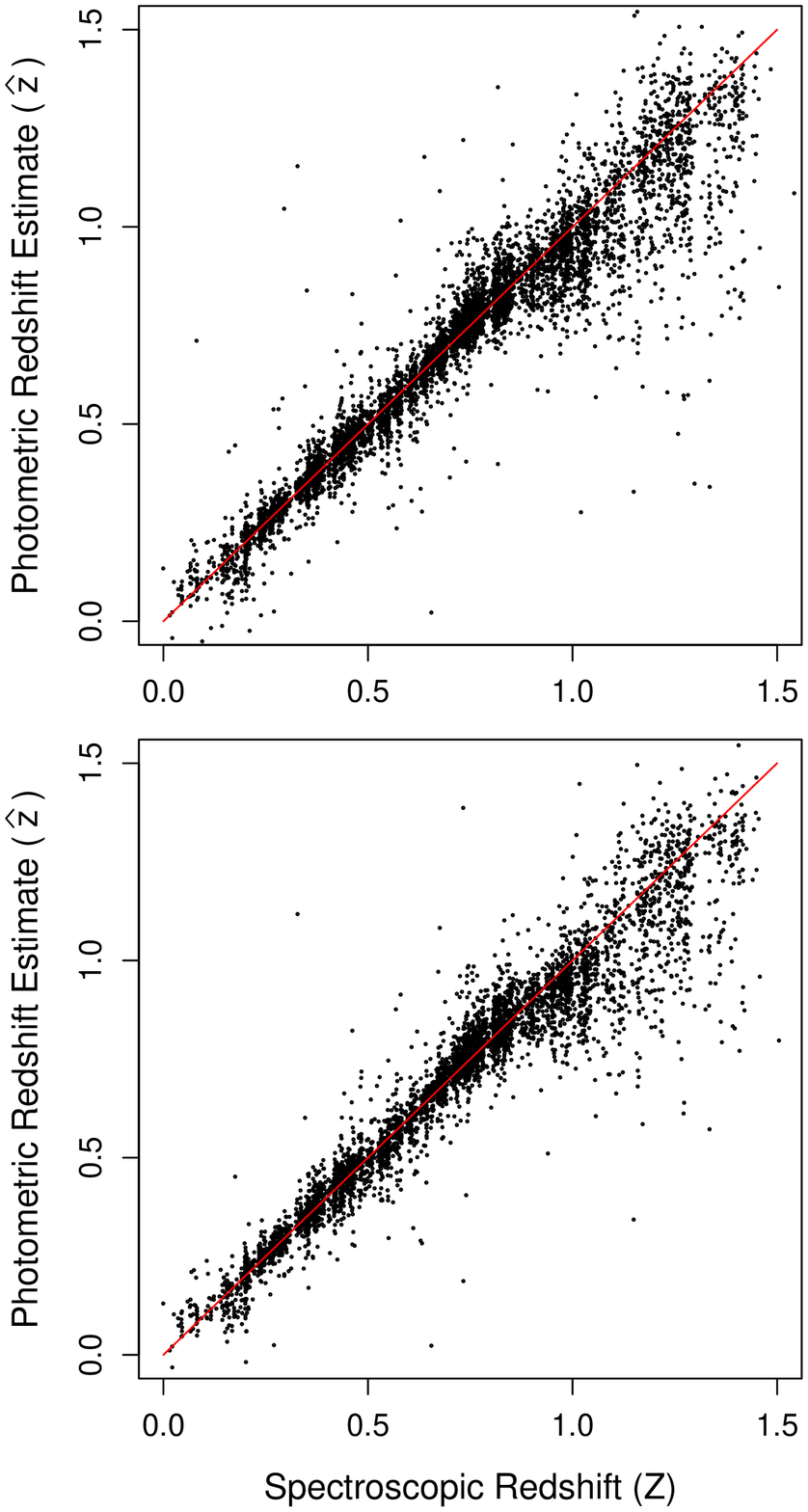}
\caption{
Top: predictions for the 6,067 objects in the DEEP2 training
set for which {\tt ZQUALITY} $>$ 3.
For these data, $(\epshat,\mhat)$ = (0.002,1050) and $\Rhat$ = 0.0539.
Bottom: same as top, for the 5,223 objects for which {\tt ZQUALITY} = 4;
$(\epshat,\mhat)$ = (0.002,850) and $\Rhat$ = 0.0507.
}
\label{fig:deep2}
\end{figure}

Application of the algorithm outlined in {\S}{\S}2.1-2.2 yields
tuning parameter estimates $(\epshat,\mhat)$ = (0.002,850) for
{\tt ZQUALITY} = 4 and (0.002,1050) for {\tt ZQUALITY} $\geq$ 3.
We display our results
in Table \ref{tab:results} and Fig. \ref{fig:deep2}; note that because
we do not apply the Nystr\"om extension here (but rather, fit to the
data directly after $(\epshat,\mhat)$ are determined), the observed
scatter is smaller than we would observe with a 
larger, Nystr\"om-extended dataset.
In both cases, we exclude 5.8\% of the objects from analysis as outliers.

\begin{figure}
\includegraphics[width=84mm]{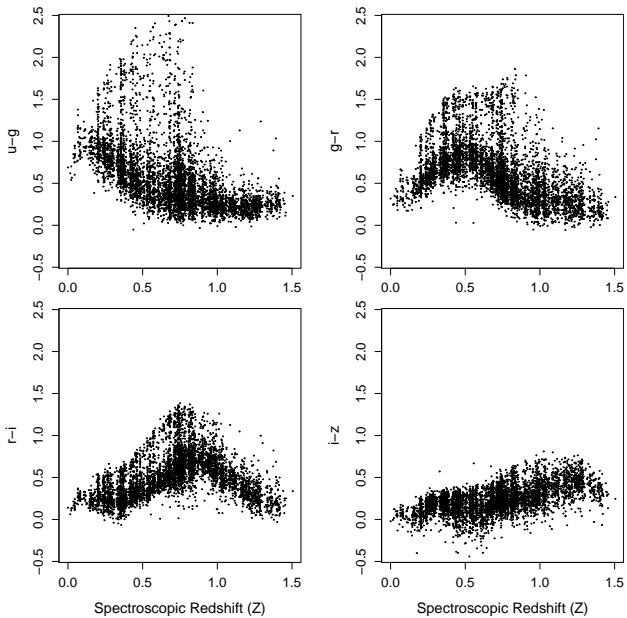}
\caption{
Observed $ugriz$ colours for the 5,223 objects in the DEEP2 training set 
for which {\tt ZQUALITY} = 4.
}
\label{fig:deep2colours}
\end{figure}

\begin{figure}
\includegraphics[width=84mm]{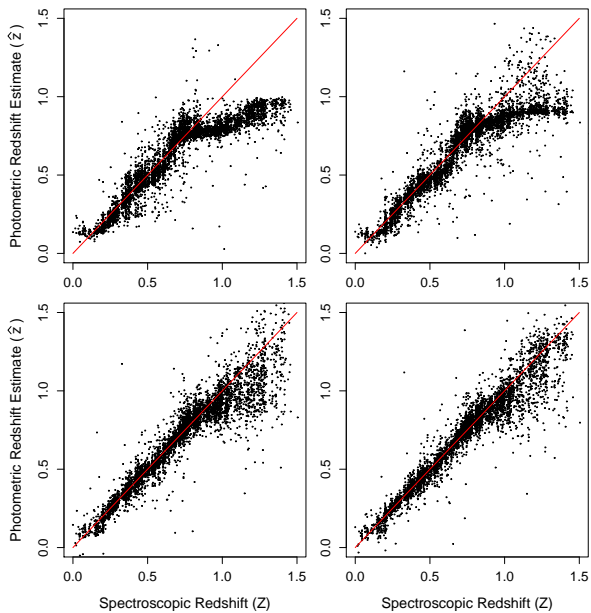}
\caption{
Predictions for the 5,223 objects in the DEEP2 training set for which
{\tt ZQUALITY} = 4, for $\epshat$ = 0.002 and $m$ = 40 (top left),
100 (top right), 400 (bottom left), and 850 ($\mhat$; bottom right).
}
\label{fig:deep2ev}
\end{figure}

In Fig. \ref{fig:deep2}, we observe that the quality of the
fits below $Z \approx$ 0.75 ($\Rhat$ = 0.038 for 
{\tt ZQUALITY} = 4) is superior
to that at higher redshifts ($\Rhat$ = 0.064).  To understand why
this is so, we examine the DEEP2 colour data (Fig. \ref{fig:deep2colours}).
Pick an object at $Z \approx 0.75$, and compute the Euclidean distance in
colour-space to a random object at any other redshift $Z \in [0,1.5]$.
This distance is a nearly constant function of $\Delta Z$; thus for
values of $\epsilon$ similar to those chosen in the SDSS analyses, there
is only a slightly lesser probability of diffusing from $Z =$ 0.75 to, e.g.,
$Z =$ 0.2 as to, e.g., $Z =$ 0.74.  To achieve accurate predictions
at $Z \approx 0.75$, $\epsilon$ must be made smaller 
(lessening the probability of large $\Delta Z$ jumps); this is what
our optimization yields.  A consequence of a smaller $\epshat$ is that the 
weighted graph of the DEEP2 objects is not fully connected
(see discussion around equation~\ref{eqn:weighted}). 
One can discern connectedness by examining the vector of eigenvalues;
for $\epshat$ = 0.002, the first $\approx$ 20 eigenvalues are all $>$ 0.95,
implying the presence of several disconnected clumps on the graph.
The most visually obvious 
manifestation of disconnectedness in the DEEP2 analysis is the
presence of a marked knee in the predictions 
at $Z \approx$ 0.75 for small values of $m$ (see Fig. \ref{fig:deep2ev}); 
the dominant
eigenvectors describe the low redshift data well, but not the high 
redshift data.  
As $m$ increases, the knee straightens out; however, because
of the bias-variance tradeoff, $m$ can only increase so much before $\Rhat$
begins to increase as well, due to increasing variance.  For
$\mhat$ = 850 ({\tt ZQUALITY} = 4) we have not yet achieved an
optimal description for the high-redshift data.  To demonstrate 
that we can achieve a better description of these data,
we split the full dataset into low- and high-redshift sets 
(at, e.g., $Z_{\rm cut}$ = 0.9) and
compute diffusion maps for each.  We find that we can achieve, e.g.,
$\Rhat \approx$ 0.035 for high-redshift data with as few as 40 eigenvectors,
while the predictions at low redshifts change only slightly.
While splitting the data yields better results for our DEEP2 sample, 
we do not propose such
splitting as part of our general diffusion map framework, for multiple reasons:
(a) it adds a tuning parameter ($Z_{\rm cut}$), (b) it complicates the
Nystr\"om extension (to which data split do we assign a new object?), and
most importantly (c) a data split can be rendered moot with the inclusion of 
new data in other bandpasses (e.g., the inclusion of near-IR data in the DEEP2
sample would mitigate the Euclidean-distance issue seen at $Z \approx 0.75$).

Concentrating on the regime $Z \la 0.75$, we find that our result
$\Rhat \approx 0.035$ with $\eta \approx$ 1.1\%
compares favorably with that of \cite{Ilbert06}, who train a template-based
photometric redshift code using 2,867 spectroscopic redshifts from the 
VIMOS VLT Deep Survey (VVDS) in the CFHTLS D1 field and obtain
$\sigma = 0.032$ and $\eta =$ 4\% (see their {\S}6.3 and fig.~14).
Our smaller catastrophic failure rate is presumably largely due to our
removal of colour-space outliers prior to analysis.  We note that
\citeauthor{Ilbert06} perform 
a similar analysis with CFHTLS $z$-band data removed,
with the result that a marked knee appears at $Z \approx$ 0.8 that is
similar to what we observe in analyzing our intrinsically bluer DEEP2 sample.
This supports the hypothesis that adding data from other bandpasses to
our DEEP2 sample will lead to a marked improvement in fit at redshifts
$Z \ga$ 1.

\section{Summary and future directions}

\label{sect:summary}

In this paper we apply an eigenmode-based framework utilizing the diffusion
map and linear regression to the problem of estimating redshifts
given SDSS and DEEP2/CFHTLS $ugriz$ photometry.  Because estimating
diffusion map coordinates via eigen-decomposition limits the size
of training sets to $\sim$ 10$^4$ objects, we implement the Nystr\"om
extension, which allows for computationally efficient estimation
of diffusion coordinates with a relatively small degradation of 
accuracy.

For our SDSS MSG sample, we train our linear regression
model on 9,749 randomly selected objects 
and via the Nystr\"om extension estimate
redshifts for another 340,989 galaxies.
Since the Nystr\"om extension is not robust to
extreme outliers, we use a nearest-neighbor algorithm to
eliminate 5$\sigma$ outliers in colour space; this eliminates 
$\approx$ 2.5\% of the MSG sample.  The loss in accuracy resulting from
use of the Nystr\"om extension is $\approx$ 2.4\% (as compared with
directly fitting the data of the training set).
For our SDSS LRG sample, we train our regression model on 9,734 objects
and via the Nystr\"om extension estimate redshifts for another 20,082,
with an outlier rate $\approx$ 3\% and a degradation of accuracy
$\approx$ 3\%.  As the DEEP2/CFHTLS sample has only $\approx$ 6,000
objects (with an outlier rate of $\approx$ 5.8\%), we do not define
a validation set to check the accuracy of predictions generated via
the Nystr\"om extension.  However, we will apply our regression model
to a test set comprised of all galaxies in CFHTLS fields D1-D4 and make
that catalog publicly available.

The observed bivariate
distributions $(\zhat,Z)$ for our SDSS datasets are 
similar to those computed by, e.g., 
\cite{Collister04} using {\tt ANNz} (specifically, for the SDSS MSG dataset)
and by \cite{Ball08} using
a numerically intensive nearest-neighbor algorithm (for both
the SDSS MSG and LRG datasets), with dispersion
on par with those techniques ($\Rhat \sim 0.02$; see \citealt{Ball08}
and references therein).  These distributions indicate that redshifts
are generally overestimated at low $Z$ and underestimated at high $Z$.
We demonstrate that 
this is a manifestation of attenuation bias, wherein measurement error
(uncertainty in the diffusion coordinates resulting from uncertainty
in the SDSS flux estimates) reduces the measured slope of the regression line.
In statistical parlance, the measured slope is not a consistent estimator
of the true slope.  In order to use photometric redshift estimates in
precision cosmology, it is vital that methods for producing consistent
estimates (i.e., mitigating the bias) be developed and implemented.
We are exploring using the
{\tt SIMEX}, or simulation-extrapolation, algorithm (e.g.,
\citealt{Carroll06}) to produce consistent estimates in a computationally
efficient manner, and we will present our results in a future publication.

For the DEEP2 data, the dominant feature in the observed bivariate
distribution, beyond attenuation bias, is a marked reduction in prediction
accuracy at redshifts $Z \ga$ 0.75.  
We demonstrate that this is due to 
a degeneracy in the colour-space manifold that would be mitigated 
with the introduction of more data from other bandpasses.  We note that 
we also can mitigate the effects of the degeneracy by splitting the
training set into low- and high-$Z$ samples, but we do not prefer this
approach because of the complexity it adds to the prediction algorithm
(through the addition of a tuning parameter $Z_{\rm cut}$ and the 
necessity of providing a quantitative measure for robustly choosing between
the two predictions we would generate for each test object).
At lower redshifts,
we find that the observed bivariate distribution $(\zhat,Z)$ compares
favorably with that derived by \cite{Ilbert06} ($\Rhat \approx$ 0.035 versus
$\sigma$ = 0.032).

Our current statistical framework yields a single photometric redshift
estimate for each object in the validation set, as opposed to
a probability distribution function (PDF) for each estimate
(cf.~\citealt{Ball08}).  This is a
valid approach for analyzing, at the very least, the galaxies of the SDSS 
sample that we consider in this work, as \citeauthor{Ball08} demonstrate
that the PDFs in the low-redshift regime are approximately normal;
we expect our single estimates to match the PDF means.  However, we would have
to alter our framework if we were to analyze quasars, for which the
PDFs are often bimodal (e.g., fig.~5 of \citeauthor{Ball08}).  Bimodality
is an indication of (near-)degeneracy in the colour-space manifold; when its
colours are perturbed, a quasar's nearest neighbor sometimes belongs to
one range of redshifts, and sometimes to a completely different range.
Within our current framework,
such a degeneracy would not affect the computation of the diffusion map,
but the subsequent
application of linear regression would yield inaccurate redshift estimates
for those quasars in the vicinity of the degeneracy.  For quasar analysis,
we would explore a variety of options, which include (a) utilizing a
different form of regression, (b) incorporating
the response variables into the construction of the diffusion map
(\citealt{Costa05}), and/or 
(c) incorporating gradient information into
diffusion map construction, such that nearby objects that lie along
the manifold have higher similarity measures.  Such schemes would mitigate
but not entirely lift the degeneracy and thus we would also have to 
quantify the relative probabilities of dual estimates.

In this work, we demonstrate the efficacy of SCA, in particular
our diffusion map framework,
for analyzing datasets for which the spectroscopic redshifts are known.
The next step is to extend our framework such that it yields accurate
photometric redshift estimates for objects in datasets where the 
spectroscopic coverage will be minimal, such as deep
sky surveys (e.g., LSST) or pointed surveys beyond $Z \approx 1$.
Even with long exposure times, the DEEP2 Galaxy Redshift Survey
is only able to 
determine secure redshifts for $\sim$ 70\% of its objects,
with about half the missed targets being star-forming galaxies at
$Z > 1.4$ that have no features in DEEP2 spectral window;
cf.~\cite{Cooper06}.  Even when spectroscopic redshifts are available
for a significant subset of these objects, it is likely that they
will be gleaned from intrinsically luminous objects whose SEDs may not
closely match those for fainter objects.  Thus it becomes imperative
to fold additional information into analyses.  \cite{Collister04},
\cite{Ball04}, and \cite{Wray08} propose
using structural properties such as surface brightness and angular radius
to obtain more
accurate redshift estimates; however, this is of limited utility at
higher redshifts.
\cite{Newman08} proposes that photometric redshifts can be calibrated 
using their correlations on the sky with objects of known redshift, 
as a function of that known redshift.  A related idea would be to 
take into account the redshifts of nearby objects on the sky in 
estimating photometric redshifts (\citealt{Kovac09});
because of the clustering of galaxies, there is a significant probability 
that two galaxies near each other on the sky are at very similar redshifts.

In a future work, we will fold additional quantities into our similarity
measure
and will determine
if photometric redshift can be estimated with sufficient accuracy so
as to fulfill their promise as a cosmological probe.

\section*{Acknowledgements}

We would like to thank both 
the referee and Larry Wasserman for helpful comments.
This work was supported by NSF grant \#0707059.
Funding for the DEEP2 survey has been provided by NSF grants AST95-09298, 
AST-0071048, AST-0071198, AST-0507428, and AST-0507483 as well as NASA LTSA 
grant NNG04GC89G.  DEEP2 data presented herein were obtained at the W. M. 
Keck Observatory, which is operated as a scientific partnership among the 
California Institute of Technology, the University of California and the 
National Aeronautics and Space Administration. The Observatory was made 
possible by the generous financial support of the W. M. Keck Foundation.
The CFHTLS data were obtained with MegaPrime/MegaCam, a joint project of 
CFHT and CEA/DAPNIA, at the Canada-France-Hawaii Telescope (CFHT) which 
is operated by the National Research Council (NRC) of Canada, the Institut 
National des Science de l'Univers of the Centre National de la Recherche 
Scientifique (CNRS) of France, and the University of Hawaii. This work is
 based in part on data products produced at TERAPIX and the Canadian 
Astronomy Data Centre as part of the Canada-France-Hawaii Telescope 
Legacy Survey, a collaborative project of NRC and CNRS.

\appendix

\section{Relevant formulae for weighted linear regression}

\label{sect:regform}

Let $\mathbfss{X}$ represent a matrix of predictors (in this
work, the matrix of diffusion coordinates $\mathbf{\Psi}$, where
each row represents the coordinates for a single object),
let $Y$ represent the vector of responses
(the estimated spectroscopic redshift values),
and let $\mathbf{\Sigma}$ represent the covariance matrix for 
$Y$, which we assume to be diagonal:
\begin{eqnarray}
\mathbf{\Sigma} = \left( \begin{array}{cccc} s_{Z_1}^2 & 0 & \cdots & 0 \\ 0 & s_{Z_2}^2 & \cdots & \vdots \\ \vdots & \vdots & \vdots & \vdots \\ 0 & \cdots & \cdots & s_{Z_n}^2 \end{array} \right) \,, \nonumber
\end{eqnarray}
Then the best-fit coefficients are
\begin{eqnarray}
{\widehat \beta} &=& \mathbfss{A} Y \nonumber \\
&=& \left(\mathbfss{X}^T \mathbf{\Sigma}^{-1} \mathbfss{X}\right)^{-1} \mathbfss{X}^T \mathbf{\Sigma}^{-1} Y \,,
\end{eqnarray}
the variance-covariance matrix for $\widehat \beta$ is
\begin{eqnarray}
\mathbb{V}({\widehat \beta}) &=& \mathbb{V}(\mathbfss{A} Y) \nonumber \\
&=& \mathbfss{A} \mathbb{V}(Y) \mathbfss{A}^T \nonumber \\
&=& \left(\mathbfss{X}^T \mathbf{\Sigma}^{-1} \mathbfss{X}\right)^{-1} \mathbfss{X}^T \mathbf{\Sigma}^{-1} \mathbb{V}(Y) \mathbf{\Sigma}^{-1} \mathbfss{X} \left(\mathbfss{X}^T \mathbf{\Sigma}^{-1} \mathbfss{X}\right)^{-1} \nonumber \\
&=& \left(\mathbfss{X}^T \mathbf{\Sigma}^{-1} \mathbfss{X}\right)^{-1} \mathbfss{X}^T \mathbf{\Sigma}^{-1} \mathbfss{X} \left(\mathbfss{X}^T \mathbf{\Sigma}^{-1} \mathbfss{X}\right)^{-1} \nonumber \\
&=& \left(\mathbfss{X}^T \mathbf{\Sigma}^{-1} \mathbfss{X}\right)^{-1}
\end{eqnarray}
and the variance-covariance matrix for $\widehat Y = \mathbfss{X} 
{\widehat \beta}$ is
\begin{eqnarray}
\mathbb{V}({\widehat Y}) &=& \mathbb{V}(\mathbfss{X} {\widehat \beta}) \nonumber \\
&=& \mathbfss{X} \mathbb{V}({\widehat \beta}) \mathbfss{X}^T \nonumber \\
&=& \mathbfss{X} \left(\mathbfss{X}^T \mathbf{\Sigma}^{-1} \mathbfss{X}\right)^{-1} \mathbfss{X}^T \label{eqn:vy} \,.
\end{eqnarray}

\section{Resampling SDSS flux measurements}

\label{sect:sample}

We assume each flux is a normal deviate with error estimated by
the Princeton/MIT data reduction pipeline.  However,
fluxes in, e.g., different SDSS magnitude bands and systems are 
correlated random variables.  In order to resample fluxes accurately,
we must take these correlations into account.  
For each object in the validation set, we have 20 flux measurements
$F$ and estimates of flux standard error $s_F$.  The covariance matrix
$\mathbf{\Sigma}$ is defined as
\begin{eqnarray}
\mathbf{\Sigma} = \left( \begin{array}{cccc} 1 & \rho_{1,2} s_{F_1} s_{F_2} & \cdots & \rho_{1,20} s_{F_1} s_{F_{20}} \\ \rho_{1,2} s_{F_1} s_{F_2} & 1 & \cdots & \vdots \\ \vdots & \vdots & \vdots & \vdots \\ \rho_{1,20} s_{F_1} s_{F_{20}} & \cdots & \cdots & 1 \end{array} \right) \,, \nonumber
\end{eqnarray}
where $\rho_{i,j}$ is the sample correlation coefficient between 
measurements $i$ and $j$ (e.g., between the PSF $u$-band 
and the Petrosian $r$-band).  We estimate $\rho_{i,j}$
using Pearson's product-moment correlation estimator
\begin{eqnarray}
\rho_{i,j} = \frac{1}{n-1} \sum_{k=1}^n \left( \frac{F_{i,k} - {\bar F}_i}{s_i}\right) \left( \frac{F_{j,k} - {\bar F}_j}{s_j} \right) \,, \nonumber
\end{eqnarray}
where $s$ is the sample standard deviation.  As expected, we find that
fluxes measured via different systems within a single magnitude band 
are strongly positively correlated ($\rho > 0.5$); also,
we find that fluxes across bands have non-negligible positive correlations,
which we attribute to the relative homogeneity of the MSG sample
(whose objects lie at relatively similar distances and display relatively
similar physical characteristics).  However,
so as not to impose this homogeneity in resampling, we set 
$\rho_{i,j}$ = 0 if indices $i$ and $j$ represent different magnitude bands.

We use the Cholesky method to decompose $\mathbf{\Sigma}$ into lower- and
upper-triangular matrices $\mathbfss{A}$ and $\mathbfss{A}^{\rm T}$.  Then we can 
compute a new vector of fluxes:
\begin{eqnarray}
F_{i}' = F_{i} + \mathbfss{A}z \nonumber \,,
\end{eqnarray}
where $z$ is a vector of standard normal deviates.

\bsp

\label{lastpage}

\end{document}